# Ansatz for the Two-Dimensional Ising Model in an External Magnetic Field

Moorad Alexanian

*Department of Physics and Physical Oceanography*
*University of North Carolina Wilmington, Wilmington, NC 28403-5606*

E-mail: alexanian@uncw.edu

**Abstract.** An ansatz applied to the two-dimensional Ising model in an external magnetic field $h$ gives rise to an exactly soluble model. The singularity in the magnetization found by Onsager does not survive the presence of the external magnetic field as found earlier by Lee and Yang in 1952. However, the singularity in the heat capacity remains even in the presence of the magnetic field. A surprising result is the presence of negative heat capacity for $h > 0$.

**Keywords:** ansatz, Ising model, Onsager, magnetization, external magnetic field, modified Ising model, phase transition, energy, positive/negative heat capacity

## 1. Introduction

The 2-dimensional (2D) Ising model in the absence of an external magnetic field was exactly solved by Onsager [1] and it was the first, exactly solved statistical mechanical model of interacting spins in a lattice. The importance of the analytic and numerical solutions of the 2D Ising model lays in establishing rigorously for the first time that thermodynamic phase transitions can indeed be characterized by mathematical singularities [1–3]. Here, we consider an ansatz applied to the 2D Ising model in an external magnetic field, which reduces the Hamiltonian to that of an exactly soluble model.

This paper is structured as follows, In Sec. 2, we present the 2D Ising model in the presence of an external magnetic field. In Sec. 3, we introduce an ansatz that gives us the canonical partition function of the modified Ising model with an external magnetic field. In Sec. 4, we calculate the magnetization versus the external magnetic field $h$ that exhibits the Onsager critical temperature for $h = 0$. In addition, we determine the magnetization versus the temperature in the presence of the external magnetic field that shows the absence of the Onsager phase transition for $h \neq 0$. In Sec. 5, we determine the energy and show that the singularity in the heat capacity remains in the presence of the external magnetic field, however, the heat capacity may assume negative values. Finally, in Sec. 6, we summarize our results.

## 2. 2D Ising model

Consider the 2D Ising model Hamiltonian with nearest-neighbor interaction strength $J$ defined over a two-dimensional, square lattice of N spins in a uniform magnetic field $h$,

$$H(\sigma_{ij}) = -J \sum_{ij} \sigma_{ij}(\sigma_{i,j+1} + \sigma_{i,j-1} + \sigma_{i+1,j} + \sigma_{i-1,j}) - h \sum_{ij} \sigma_{ij}, \qquad (1)$$

with $\sigma_{ij} = \pm 1$.

The seminal solution of Onsager [1] in the absence of a magnetic field ($h = 0$) for the canonical partition function [3] is

$$\mathscr{Z} = \lambda^N, \qquad (2)$$

where

$$\ln \lambda = \ln(2\cosh 2\beta J) + \frac{1}{\pi}\int_0^{\frac{\pi}{2}} dx \ln\left[\frac{1}{2}\{1+(1-k^2\sin^2 x)^{1/2}\}\right], \quad (3)$$

$$k = \frac{2\sinh 2\beta J}{\cosh^2 2\beta J}, \quad \text{and} \quad \beta = 1/k_B T. \quad (4)$$

## 3. Ansatz

Consider the following ansatz for the spin interaction terms with the its neighbors in (1)

$$\sigma_{i\pm 1,j} = \alpha\left(\frac{1\pm\sigma_{ij}}{2}\right) \quad \text{and} \quad \sigma_{i,j\pm 1} = \alpha\left(\frac{1\pm\sigma_{ij}}{2}\right), \quad (5)$$

where α is a real function of βJ.

Hamiltonian (1) becomes

$$H_{ans}(\sigma_{ij}) = -\sum_{ij}(2J\alpha + h)\sigma_{ij} \quad (6)$$

with the canonical partition function

$$\mathcal{Z}_{ans} = \left[2\cosh\beta(2\alpha J + h)\right]^N. \quad (7)$$

We determine the function α by requiring that the canonical partition function (7) of the ansatz Hamiltonian (6) for $h = 0$ is equal to the Onsager canonical partition function and so

$$\ln\left(\frac{\cosh 2\alpha\beta J}{\cosh 2\beta J}\right) = \frac{1}{\pi}\int_0^{\frac{\pi}{2}} dx \ln\left[\frac{1}{2}\{1+(1-k^2\sin^2 x)^{1/2}\}\right] \equiv f(k), \quad (8)$$

thus

$$\alpha(T) = \pm\frac{T}{2J}\text{arcCosh}\left(\cosh(2J/T)e^{f(k)}\right), \quad (9)$$

where the ± sign is associated with the sign of h. Accordingly, we recover all the Onsager results for thermodynamic observables with the aid of α(T) determined by (9) and using the ansatz (7) will allow us to extend all Onsager results for $h = 0$ to any non-zero magnetic field h.

## 4. Magnetization

The magnetization M follows from the canonical ensemble $\mathcal{Z}$

$$M = \left(\frac{\partial\ln\mathcal{Z}}{\partial h}\right)_\beta = N\tanh\left(\beta(2\alpha J + h)\right) \quad (10)$$

with the aid of (7). We shall use the exact zero-field magnetization found by Onsager to aid us in finding the magnetization for arbitrary values of the external magnetic field h. Besides giving the exact solution of the 2D Ising model in zero external magnetic field, Onsager also gave, without derivation, the expression for the magnetization for $T \leq T_c$

$$\frac{M}{N} = \left[1 - \sinh^{-4}(2\beta J)\right]^{1/8}. \quad (11)$$

where

$$k_B T_c = \frac{2J}{\ln(1+\sqrt{2})} = 2.269\cdots J. \qquad (12)$$

We then obtain, with the aid of the canonical partition function (7), that

$$\alpha_m(T) = \pm\frac{T}{2}\operatorname{arctanh}\left(\frac{\sinh^4(2J/T)-1}{\sinh^4(2J/T)}\right)^{1/8}, \qquad (13)$$

for $T \leq T_c$, where the $\pm$ sign is associated with the sign of $h$, and so that the magnetization per particle is given by

$$M_{\leq}(T,h) = \tanh\left(\beta\big(2J\alpha_m(T)\operatorname{sign}(h) + h\big)\right) \qquad \text{for} \qquad T \leq T_c, \qquad (14)$$

which is an odd function of $h$ and where sign($x$) = 1 for $x \geq 0$, and −1 for $x < 0$. Note the discontinuity in the magnetization versus the magnetic field (14) at $h$ = 0. Fig.1(a) shows the behavior of the magnetization versus external magnetic field $h$ for $T \leq T_c$, where we have set $J = 1$ and $k_B = 1$, i.e., expressing the energy and the temperature in units of $J$.

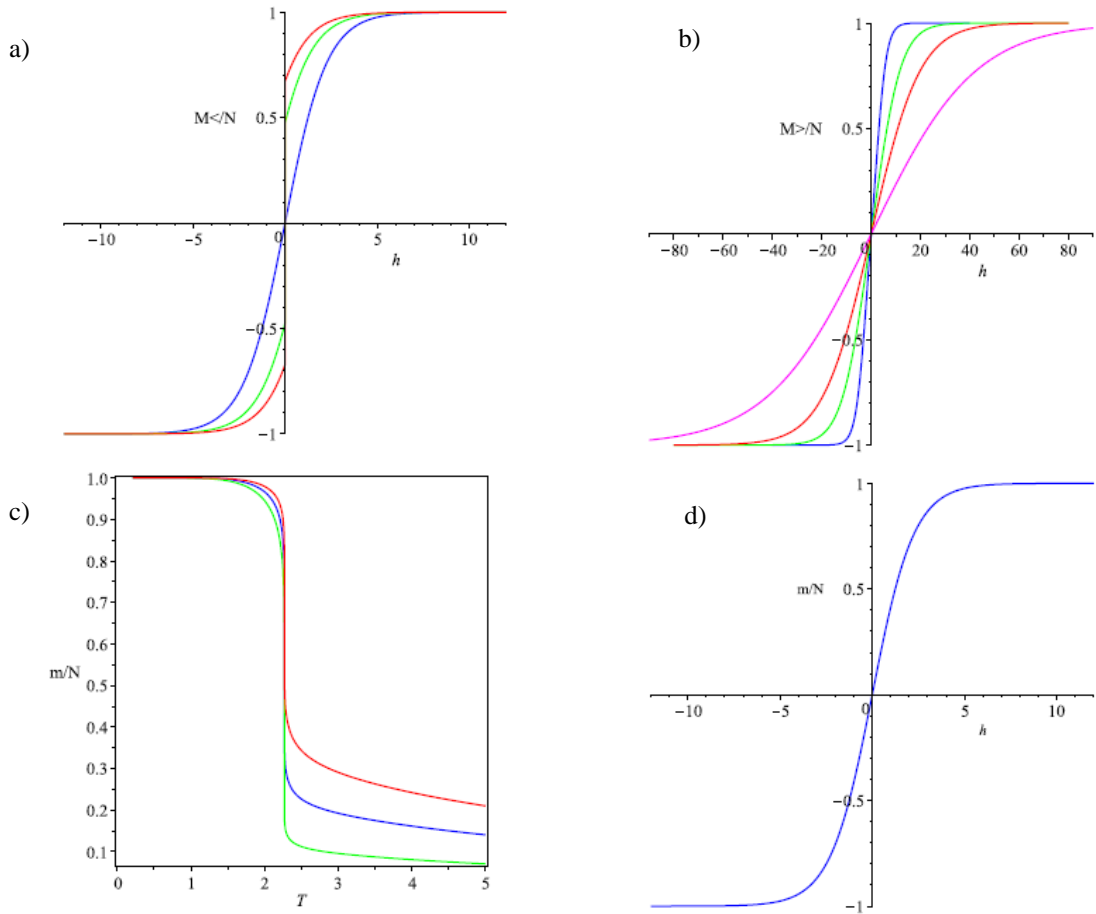

**Fig. 1.** (a) Magnetization for $T \leq T_c$, $T = T_c = 2/\ln(1 +\sqrt{2}) = 2.26918...$(blue), $T$ = 2.268 (green), $T$ = 2.25 (red). (b) Magnetization for $T > T_c$, $T$ = 4 (blue), $T$ = 10 (green), $T$ = 18 (red), $T$ = 40 (magenta). (c) Magnetization per particle, $m/N$, versus temperature $T$ for differing values of magnetic field $h$, $h$ = 0.5 (green), $h$ = 1 (blue),  $h$ = 1.5 (red). (d) Magnetization per particle, $m/N$, versus magnetic field $h$ evaluated at the critical temperature $T_c$. Same as blue plot in Fig.1a.

Unfortunately, we cannot extend (14) for values $T \geq T_c$ since $\alpha_m$ assumes complex values. We consider, instead, the real value of the complex function $\alpha_m(T)$ for $T \geq T_c$. Accordingly,

$$M_\geq(T,h) = \frac{1}{2}\tanh\left(\frac{[\alpha_m(T)+\alpha_m^*(T)]\text{sgn}(h)+h}{T}\right) - \frac{1}{2}\tanh\left(\frac{[\alpha_m(T)+\alpha_m^*(T)]\text{sgn}(h)-h}{T}\right) \quad \text{for} \quad T \geq T_c, \tag{15}$$

where $\text{sgn}(x) = 1$ for $x > 0$, 0 for $x = 0$, and -1 for $x < 0$. Fig. 1(b) shows the magnetization versus $h$ for $T \geq T_c$.

It is interesting that at $T = T_c$ one has that the magnetization does not vanish for $h \neq 0$ and so

$$M(T_c, h) = \tanh(h/T_c) \tag{16}$$

from (14)-(15) since $\alpha_m(T_c) = 0$. Therefore, when the external magnetic field $h$ is non-zero, the magnetic susceptibility per particle, $\chi = \partial M/\partial h$, versus $T$ is now continuous at $T_c$. There is no thermodynamic phase transition when $h \neq 0$ as first proved by Lee and Yang in 1952 [4].

One can illustrate the absence of the phase transition by considering

$$M(T,h) = M_\leq(T,h)H(T_c - T) + M_\geq(T,h)H(T - T_c), \quad \text{for} \quad T \geq 0, \tag{17}$$

where $H(x)$ is the Heaviside function with $H(x) = 1$ for $x > 0$, 0 for $x < 0$, and 1/2 for $x = 0$, viz., $H(x) = 1/2 \cdot [1 + \text{sgn}(x)]$. Fig.1(c) shows the behavior of the magnetization versus the temperature for differing values of the magnetic field $h$ as given by (17). Note the absence of the singularity at $T = T_c$. In Fig.1(d), the plot shows of the magnetization at the critical temperature $T_c$ versus the external magnetic field $h$.

## 5. Energy

The energy E follows from the canonical partition function (7) and (8),

$$E = -\left(\frac{\partial \ln \mathscr{Z}}{\partial \beta}\right)_h = -N\frac{\tanh(2\alpha\beta + h\beta)}{\tanh(2\alpha\beta)}\left(2\tanh(2\beta) + \frac{df(k)}{d\beta} + h\tanh(2\alpha\beta)\right). \tag{18}$$

The integral $f(k)$ in (8) cannon be expressed in terms of know functions. However, its derivative gives rise to an elliptic function, one has that

$$\frac{df(k)}{d\beta} = \frac{df(k)}{dk}\frac{dk}{d\beta} = 4\left(\frac{2-\cosh^2(2\beta)}{\cosh^3(2\beta)}\right)\left(\frac{\pi - 2K(k)}{2\pi k}\right), \tag{19}$$

where $K(k)$ is the complete elliptic integral of the first kind

$$K(k) = \int_0^1 \frac{dx}{\sqrt{(1-x^2)(1-k^2x^2)}}, \tag{20}$$

$$\frac{dk}{d\beta} = \frac{4(2-\cosh^2(2\beta))}{\cosh^3(2\beta)}, \quad \text{and} \quad \frac{df(k)}{dk} = \frac{\pi - 2K(k)}{2\pi k}. \tag{21}$$

The energy per particle $e(\beta,h) = E/N$ follow from (18)-(21) and so where $e(\beta)$ is the Onsager result for the energy per particle, viz., at $h = 0$,

$$e(\beta) = -2\tanh(2\beta) - \frac{1}{\pi}\frac{(\pi - 2K(k))(2-\cosh^2(2\beta))}{\sinh(2\beta)\cosh(2\beta)}. \tag{23}$$

$$e(\beta, h) = \frac{\tanh(\beta(2\alpha + h))}{\tanh(2\alpha\beta)}(e(\beta) - h\tanh(2\alpha\beta)), \tag{22}$$

Fig.2(a) shows the energy per particle $E/N$ versus the temperature $T$ for different values of the external magnetic field $h$. The energy per particles $E/N \to -2 - h$ as $T \to 0$. Note the heat capacity $C_V = (\partial E/\partial T)_V$ for $h > 0$, where the heat capacity assumes negative values.

On considering the heat capacity $C_V$ from (22), we encounter the general form $\partial \tanh(2\alpha\beta)/\partial\beta$ which leads to

$$\tanh(2\alpha\beta)(2\alpha + 2\beta\frac{d\alpha}{d\beta}) = 2\tanh(2\beta) + \frac{df(k)}{d\beta} \tag{24}$$

with the aid of (9). One obtains then with the aid of (19) and (21) that

$$e(\beta) = -2(\alpha + \beta\frac{d\alpha}{d\beta})\tanh(2\alpha\beta). \tag{25}$$

The left hand side of (25) is a regular function of $\beta$ and so is then the right hand side of (25). Accordingly, the singularity in the specific heat $C_V$ at the critical temperature $T_c$ follows only from the term with $de(\beta)/d\beta$ in (22) as in the Onsager case and so the singularity of the specific heat at the critical temperature $T_c$ remains even in the presence of the external magnetic field $h$. See Fig.2(b). This is quite different from the singularity in the magnetization that gives rise to a phase transition when $h = 0$ that does not prevail in the presence of a non-zero external magnetic field $h$.

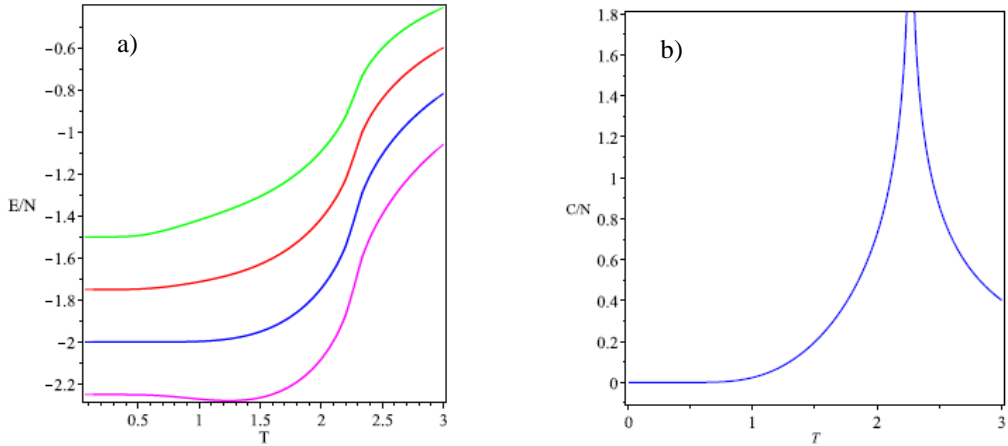

**Fig. 2.** (a) Energy per particle $E/N$ versus the temperature $T$ for different values of the external magnetic field $h$, $h = 0$ (blue), $h = -0.5$ (green), $h = -0.25$ (red), and $h = 0.25$ (magenta). (b) Heat capacity versus temperature with external magnetic field $h = 0$.

The transition from a positive to a negative value for the specific heat occurs at $h = 0$. Now it follows from (9) that

$$\alpha(T) = \pm\left(1 - \frac{T}{2}e^{-4/T} + O(e^{-8/T})\right), \tag{26}$$

where the $\pm$ sign is associated with the sign of $h$, Also,

$$K(k) = (1/2)\pi + (1/8)\pi k^2 + (9/128)\pi k^4 + O(k^6). \tag{28}$$

and

The energy per particle (22) assumes quite different values for positive or negative values of $h$. Consider first the case $h > 0$. We have that $e^{\beta h} \gg 1$ for $\beta \gg 1$ and so

$$e(\beta, h) = -2 - h - (2h + 8)(\exp(-2\beta))^2 - (2h + 8)(\exp(-2\beta))^4 + O((\exp(-2\beta))^6) \tag{29}$$

with heat capacity

$$C_V = -4(2h+8)T^{-2}(\exp(-2/T))^2 - 8(2h+8)T^{-2}(\exp(-2/T))^4 + O((\exp(-2/T))^6), \qquad (30)$$

with an essential singularity at $T = 0$. Note the negative value for the heat capacity for $h > 0$. See Fig.2(a).

The case with $h < 0$ is quite distinct. Now $e^{\beta h} \ll 1$ for $\beta \gg 1$ and so (22) becomes

$$e(\beta, h) = -2 + h - 2(\exp(-2\beta))^2 h + (8 + 2h)(\exp(-2\beta))^4 + O((\exp(-2\beta))^6) \qquad (31)$$

with heat capacity

$$C_V = -8T^{-2}(\exp(-2/T))^2 h + 8(8+2h)T^{-2}(\exp(-2/T))^4 + O((\exp(-2/T))^6), \qquad (32)$$

which is positive, since $h < 0$ and possesses an essential singularity at $T = 0$. See Fig.2(a).

Thermodynamics dictates that the specific heat of a system is strictly non-negative. However, negative heat capacity may arise in low-dimensional/nano quantum oscillators [5]. In addition, in finite classical systems there are well-known theoretical and experimental cases where this rule is violated, in particular finite atomic clusters. [6] Negative heat capacity can also occur in finite quantum systems [7].

## 6. Conclusions

We have introduced an ansatz in the 2D Ising model with an external magnetic field $h$ that allows us to evaluate exactly the magnetization and the energy for arbitrary values of $h$ and temperature $T$. We find that for $h \neq 0$, there is no thermodynamic phase transition in the magnetization, which was first proved by Lee and Yang. However, the presence of a nonzero $h$ does not remove the singularity at the critical temperature $T_c$ in the heat capacity. In addition, there is a sort of phase transition at $h = 0$ since the heat capacity assumes negative values for $h > 0$.